\documentclass[12pt]{iopart}

\usepackage{graphicx}

\begin{document}

\title[Optical and magneto-optical interactions in Co-doped CeO$_2$ thin films]{Optical and magneto-optical interactions in Co-doped CeO$_2$ thin films prepared by pulsed laser deposition}

\author{Martin Zahradn\'{i}k$^{1,2}$, Miroslav Ku\v{c}era$^1$, Roman Anto\v{s}$^1$, Martin Veis$^{1,*}$, Jan Mistr\'{i}k$^3$, Lei Bi$^4$, Hyun-Suk Kim$^5$, and Caroline A. Ross$^6$}

\address{$^1$ Charles University, Faculty of Mathematics and Physics, Ke Karlovu 3, 12116 Prague 2, Czech Republic}
\address{$^2$ ELI Beamlines Facility, The Extreme Light Infrastructure ERIC, Za Radnic\'{i} 835, 25241 Doln\'{i} B\v{r}e\v{z}any, Czech Republic}
\address{$^3$ University of Pardubice, Department of Physics, 53210 Pardubice, Czech Republic}
\address{$^4$ University of Electronic Science and Technology of China, Chengdu, China}
\address{$^5$ Materials Science \& Engineering, Chungnam National University, 99 Daehak-ro, Yuseong-gu, Daejeon 305-764, Republic of Korea}
\address{$^6$ Department of Materials Science and Engineering, Massachusetts Institute of Technology, Cambridge, USA}
\address{$^*$ Author to whom any correspondence should be addressed.}
\ead{martin.veis@matfyz.cuni.cz}

\begin{abstract}
Magnetically doped CeO$_2$ is a dilute magnetic semiconductor, promising for various applications in photonics, but the origin of its ferromagnetic properties is not fully understood. Here, thin films of Ce$_{1-x}$Co$_x$O$_{2-\delta}$ prepared by pulsed laser deposition on MgO ($x=0.05$ and $0.10$) and oxidized Si ($x=0.20$) substrates were systematically studied by spectroscopic ellipsometry and magneto-optical spectroscopy. Both diagonal and off-diagonal permittivity-tensor elements were obtained. Diagonal spectra revealed two optical transitions between oxygen and cerium states. Off-diagonal spectra revealed two paramagnetic transitions involving cobalt ions, from which an essential influence of cobalt doping on resulting ferromagnetic properties of CeO$_2$ was inferred. The full permittivity-tensor spectra are provided for further use in prospective modelling of magneto-optical device concepts.
\end{abstract}

\vspace{2pc}

\section{Introduction}

Magnetically doped CeO$_2$ is a promising candidate for many applications in photonic technologies that use magneto-optical materials, such as magneto-optical isolators \cite{Bi2011}, magneto-photonic crystals \cite{Inoue2006a,Inoue2006b}, magneto-optical spatial light modulators \cite{Iwasaki2006,Iwasaki2008}, or magneto-plasmonic crystals \cite{Belotelov2011,Pohl2013}. These devices require materials with tunable optical, magnetic, and magneto-optical properties, which can be adjusted via the incorporation of magnetic ions into the non-magnetic lattice. Magnetically doped CeO$_2$ is a dilute magnetic semiconductor with high Curie temperature and a range of physical properties that depend on the Co doping \cite{Vodungbo2007,Fernandes2007,Tiwari2006,Bi2008}. Its surface chemistry makes it useful as a catalyst in advanced chemi-resistive sensors \cite{Baek2024}. Detailed knowledge of its optical and magneto-optical properties, i.e. the complete permittivity tensor, is necessary to understand the magneto-optical performance, and to successfully adopt this material in proposed device concepts.

Room-temperature (RT) ferromagnetism in pure CeO$_2$ has been observed and is typically explained in terms of lattice defects, such as oxygen interstitials and antisites \cite{ElHachimi2014}, cerium vacancies \cite{ElHachimi2014,Fernandes2009,Lu2012}, and especially oxygen vacancies \cite{Vodungbo2007,Fernandes2007,Fernandes2009,Chen2010,Jarlborg2014}. Oxygen vacancy-mediated magnetism has been inferred from the change from ferromagnetic to paramagnetic behavior after oxygen annealing \cite{Wen2007,Singhal2010}, or the enhanced ferromagnetism of CeO$_2$ after hydrogenation that is directly linked to an increased amount of oxygen vacancies \cite{Li2008,Singhal2012}. On the other hand, other authors have found no correlation of the ferromagnetism in CeO$_2$ with oxygen vacancies \cite{ElHachimi2014,Liu2008}. Beside pure CeO$_2$, the role of oxygen vacancies has been extensively discussed also in magnetically doped CeO$_2$ \cite{Murugan2016,Ackland2018,Yang2023,Chahal2023}. Several microscopic models have been proposed to explain the magnetic moment originating from specific defects, including conventional ferromagnetism, charge-transfer ferromagnetism \cite{Coey2008}, and giant orbital paramagnetism \cite{Coey2016}.

Here, we present a systematic study of optical and magneto-optical properties of Co doped CeO$_2$ thin films in a spectral range from near infrared (IR) to ultraviolet (UV). We provide complete permittivity tensor of this material, and propose an electronic model describing its RT ferromagnetism.

\section{Experimental details}

The films of Ce$_{1-x}$Co$_x$O$_{2-\delta}$ were prepared by pulsed laser deposition on substrates of MgO(100), from targets with $x=0.05$ (Co5) and $0.10$ (Co10), and naturally oxidized Si(100), $x=0.20$ (Co20). The deposition process was carried out in a vacuum of $10^{-6}$ Torr base pressure with the substrate heated at 700 $^{\circ}$C. X-ray diffraction (XRD) revealed polycrystalline films. Their thicknesses were determined by profilometer, and they were 753, 262, and 391 nm for Co5, Co10, and Co20, respectively. Wavelength dispersive spectroscopy revealed actual cobalt concentrations to be 2 \% (Co5) and 6 \% (Co10). More details on structural and magnetic characterization of Co5 and Co10 can be found in earlier reports \cite{Bi2008}.

Spectroscopic ellipsometry was performed on a high precision Woollam VASE ellipsometer. The measurements were carried out at three different angles of incidence (65$^{\circ}$, 70$^{\circ}$, and 75$^{\circ}$) in a spectral range from 1 to 5 eV. The same instrument was used to acquire spectra of optical reflectivity and transmission. All the data were analyzed simultaneously assuming a model structure of an MgO substrate and a homogeneous Ce$_{1-x}$Co$_x$O$_{2-\delta}$ layer covered with a rough surface layer. The optical parameters of the Ce$_{1-x}$Co$_x$O$_{2-\delta}$ layers, their thicknesses, and surface roughness were adjusted by the least squares method.

Faraday magneto-optical (MO) spectroscopy was performed using azimuth modulation technique with rotating polarizer and synchronous detection \cite{Kucera1985}. The Faraday rotation and magnetic circular dichroism (MCD) spectra were measured in a photon energy range from 0.6 to 3.8 eV. The obtained Faraday rotation spectra were corrected for the MgO substrate contribution. The low temperature (LT) measurements were performed using a liquid nitrogen cooled cryostat, maintaining the sample temperature at approximately 80 K. The applied magnetic field of 0.6 T was sufficient for sample saturation. Kerr MO spectroscopy was performed using generalized MO ellipsometry with rotating analyzer. The measurements were carried out at room temperature, in polar configuration, and an applied magnetic field of 1 T.

\section{Results and discussion}

\begin{figure}
	\centering
	\includegraphics[scale=0.75]{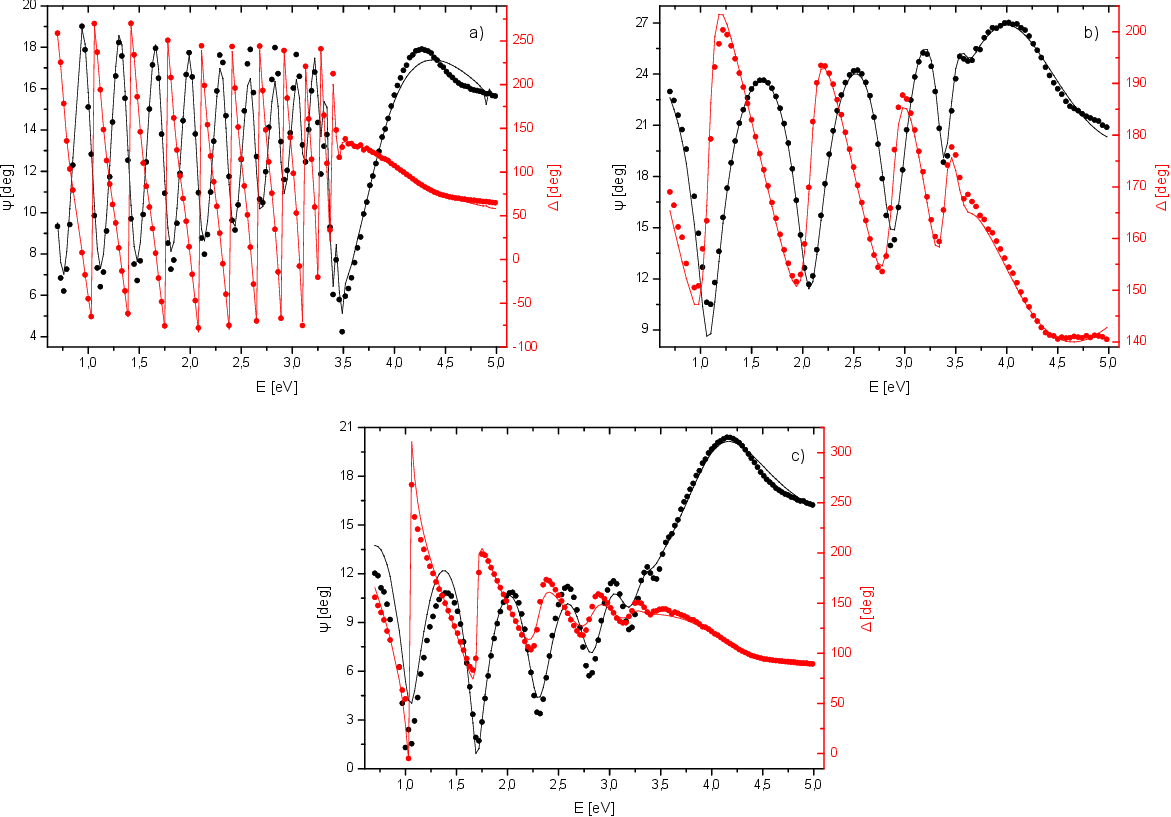}
	\caption{Experimental (symbols) and modelled (lines) spectra of ellipsometric parameters $\psi$ and $\delta$ for Ce$_{1-x}$Co$_x$O$_{2-\delta}$ films with a) $x=0.05$ at 65$^{\circ}$ angle of incidence, b) $x=0.10$ at 70$^{\circ}$, and c) $x=0.20$ at 75$^{\circ}$.}
	\label{fig:PsiDelta}
\end{figure}

The experimental spectra of ellipsometric parameters $\psi$ and $\delta$ are displayed in Figure~\ref{fig:PsiDelta} together with theoretical fits used to obtain the spectral dependencies of the real and imaginary parts of the diagonal element of the permittivity tensor. For each sample, we show an example fit at one angle of incidence, while the resulting optical parameters are acquired from a combined fit at all three angles of incidence.

\begin{figure}
	\centering
	\includegraphics[scale=0.5]{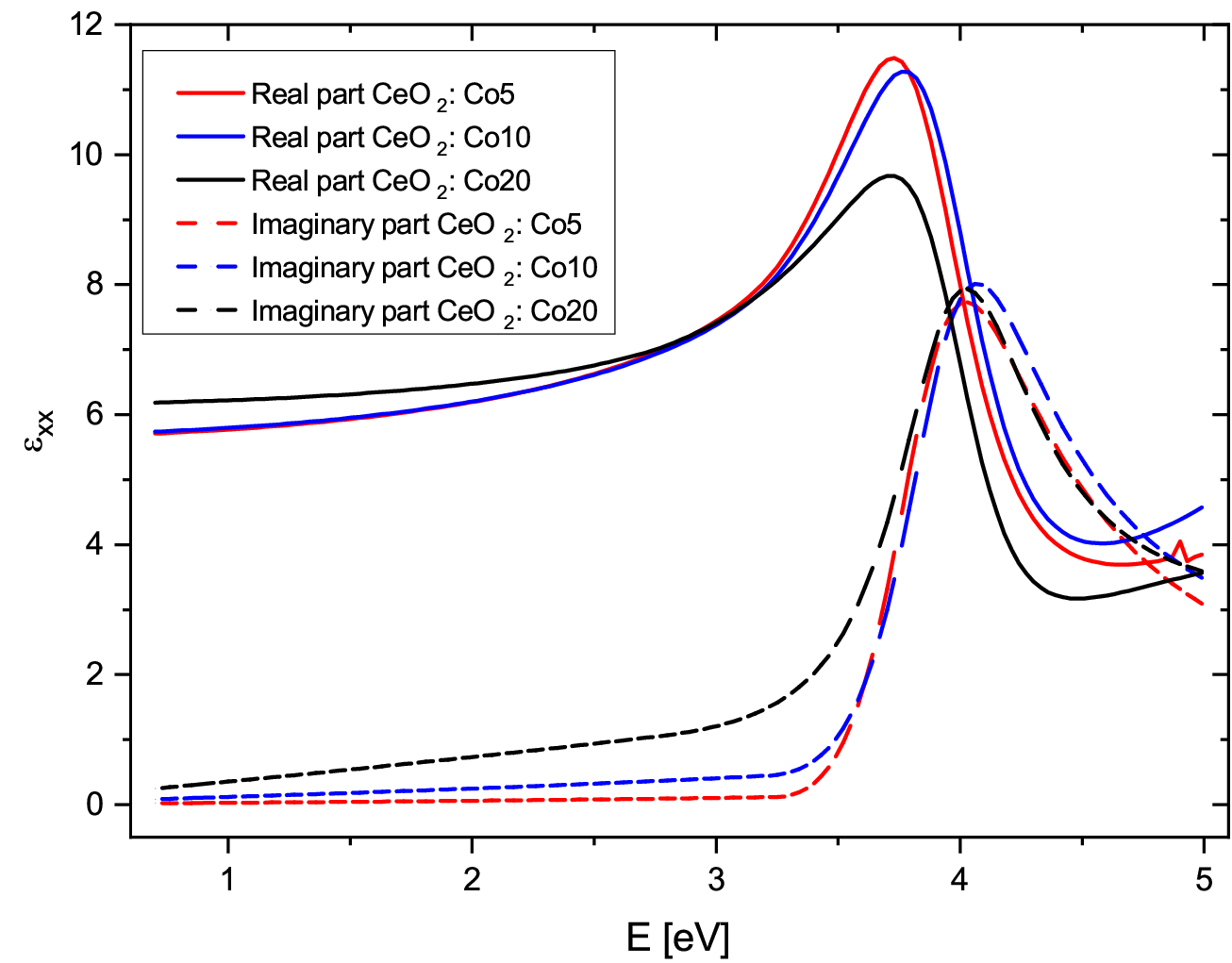}
	\caption{Spectra of the real and imaginary parts of the diagonal elements of Ce$_{1-x}$Co$_x$O$_{2-\delta}$ films with $x=0.05, 0.10$ and $0.20$, calculated from ellipsometric data.}
	\label{fig:DiagElements}
\end{figure}

The spectra of both the real and imaginary parts of the diagonal element of the permittivity tensor are shown in Figure~\ref{fig:DiagElements} for all the samples. The spectral behaviour is similar to results obtained on undoped CeO$_2$ films \cite{Barreca2003}. The theoretical fits revealed two optical transitions at around 3.9 and 8.1 eV, which are both assumed to originate charge transfer, the first from O $2p$ to Ce $4f$ states and the latter from O $2p$ to Ce $5d$ states \cite{Barreca2003}. The values of optical bandgap energies were estimated as 3.25 eV (Co5) and 3.21 eV (Co10), which is also in a good agreement with $3.23\pm0.05$ eV obtained on undoped CeO$_2$ \cite{Chiu2010}. On the other hand, the imaginary part of the diagonal element of the permittivity tensor shows noticeably larger amplitude compared to that of undoped CeO$_2$ films, which indicates a stronger absorption in cobalt doped samples. Additionally, the amplitude increases with increasing cobalt content throughout almost the whole investigated spectral range up to around 4 eV. Below the absorption edge, this may be explained in terms of midgap defects arising from oxygen vacancies or cobalt states. Oxygen vacancies are expected to change the valence of the cerium ions from Ce$^{4+}$ to Ce$^{3+}$ via localized Ce $4f$ electrons \cite{Castleton2007}. They act as recombination centres, which leads to enhanced optical absorption, as already suggested in earlier reports \cite{Veis2014}. Since our doping levels up to $x=0.20$ are expected not to change the fluorite crystal structure of CeO$_2$, increasing cobalt content leads to enhanced defect concentration, which is then linked to increased absorption. However, we do not expect the oxygen vacancy concentration to be particularly high, since the bandgap energy did not fall below 3.2 eV. Earlier reports showed bandgap narrowing related to doping-induced oxygen vacancies as large as ~300 meV \cite{Ansari2014}, while our films do not show significant bandgap narrowing with increased Co concentration.

\begin{figure}[b]
	\centering
	\includegraphics[scale=0.5]{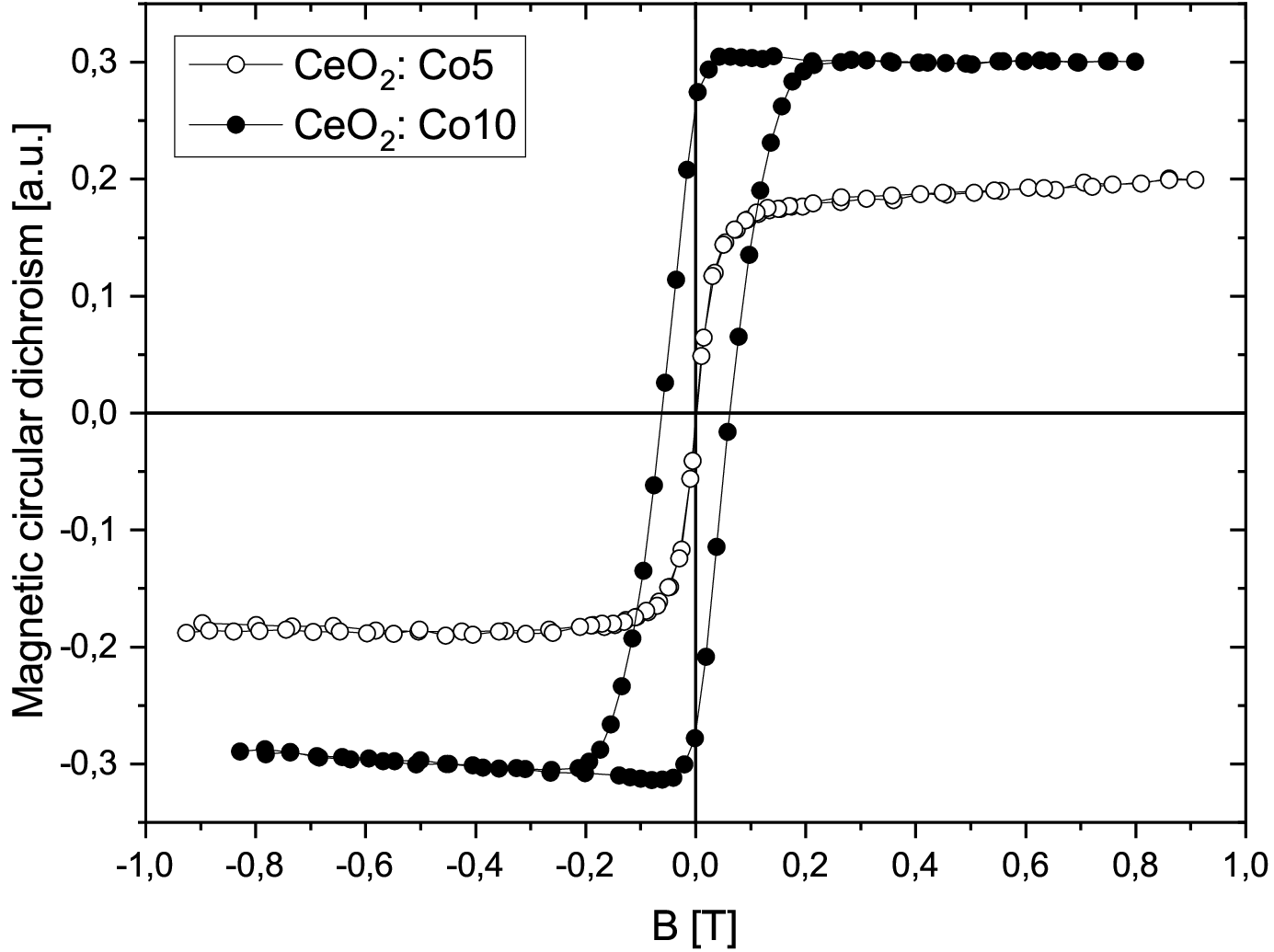}
	\caption{Hysteresis loops of magnetic circular dichroism of Ce$_{1-x}$Co$_x$O$_{2-\delta}$ films with $x=0.05$ and $0.10$, measured at room temperature with magnetic field perpendicular to the sample surface and incident photon energy of 3.3 eV.}
	\label{fig:MCDloops}
\end{figure}

Figure~\ref{fig:MCDloops} shows MCD hysteresis loops for Co5 and Co10 samples measured at room temperature with magnetic field perpendicular to the sample surface and incident photon energy of 3.3 eV. We can clearly see an RT ferromagnetic behaviour and a dependence of MO properties on cobalt content. Saturation magnetization, coercivity, and magnetic remanence increase with increasing cobalt content. The Co5 sample exhibits almost zero coercivity, which is common for undoped CeO$_2$ \cite{Sundaresan2006}. However, the Co10 sample shows hysteretic behaviour typical for magnetic anisotropy with out-of-plane easy axis, with estimated coercivity around 0.06 T. When measured in a magnetic field parallel to the sample surface, the in-plane magnetization cannot be saturated even at 1 T. Similar behaviour was already reported for thin films of cobalt-doped CeO$_2$, where the perpendicular magnetic anisotropy was explained by the doping level and the compressive strain in the film \cite{Vodungbo2008}. 

\begin{figure}
	\centering
	\includegraphics[scale=0.5]{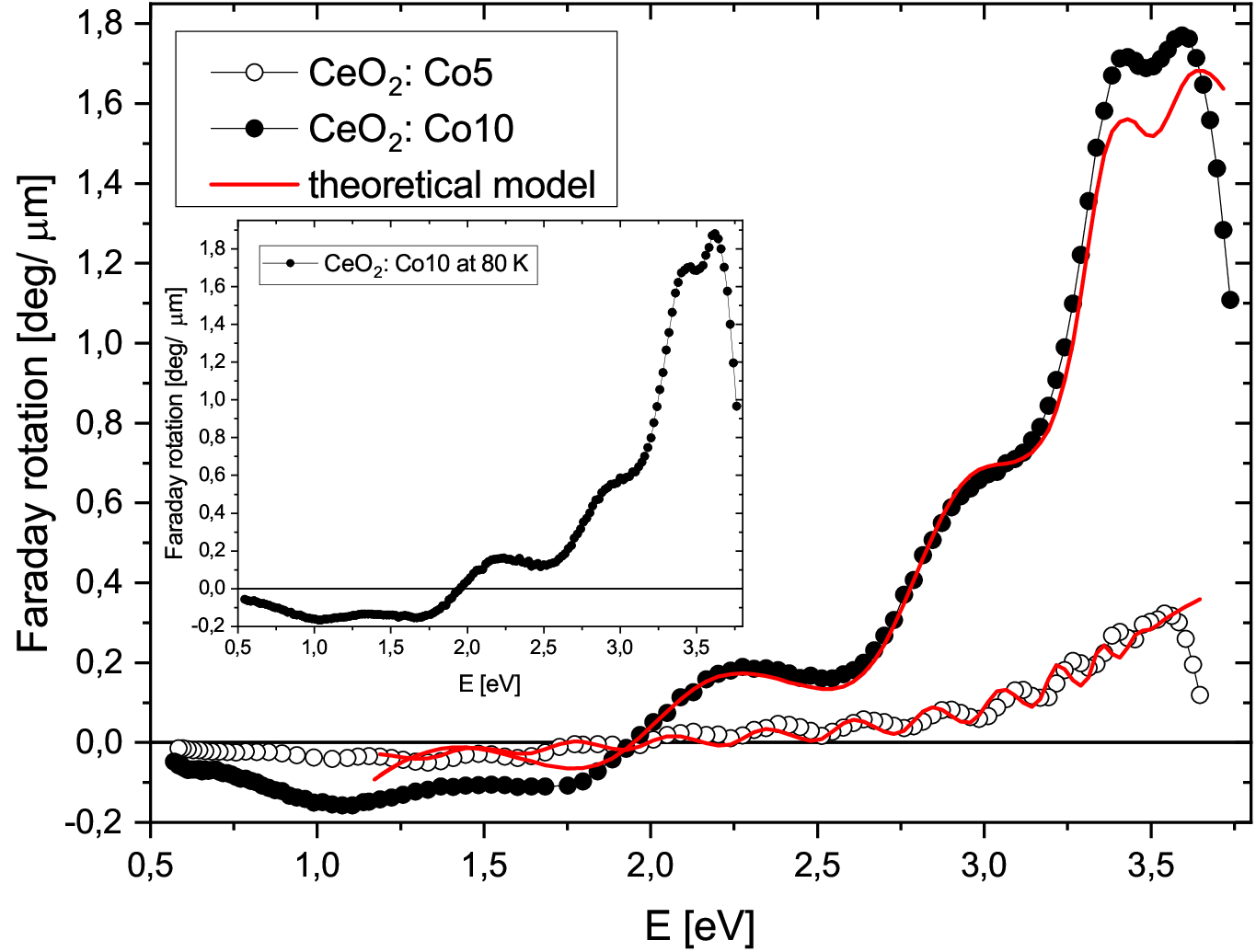}
	\caption{Experimental and modelled Faraday rotation spectra of Ce$_{1-x}$Co$_x$O$_{2-\delta}$ films with $x=0.05$ and $0.10$, measured at room temperature. The experimental Faraday rotation data are corrected for MgO substrate contribution. Theoretical spectra are modelled by two paramagnetic transitions. The inset shows experimental Faraday rotation spectra of the Ce$_{0.90}$Co$_{0.10}$O$_{2-\delta}$ film measured at 80 K.}
	\label{fig:FRspectra}
\end{figure}

RT FR spectra are shown in Figure~\ref{fig:FRspectra} together with modelled FR for Co5 and Co10 samples. Interference fringes from multiple reflections inside the films are visible for both samples. Two main spectral features are centred near 1.1 and 3.5 eV, exhibiting clear dependence of their amplitude on cobalt content. The amplitude difference is mostly enhanced near the absorption edge, which correlates with the observed increased absorption (see Figure~\ref{fig:DiagElements}) attributed to midgap defects. However, the absorption increase between Co5 and Co10 only amounts to few percent, while the FR at 3.5 eV is around five times larger for Co10 compared to Co5. This yields a MO figure of merit, defined as the ratio of FR to optical loss, more than four times higher in case of Co10. Such an increase indicates that the mechanisms governing the MO properties of the samples might differ from the midgap defects responsible for enhanced absorption, involving more likely the cobalt ions.

\begin{figure}[b]
	\centering
	\includegraphics[scale=0.5]{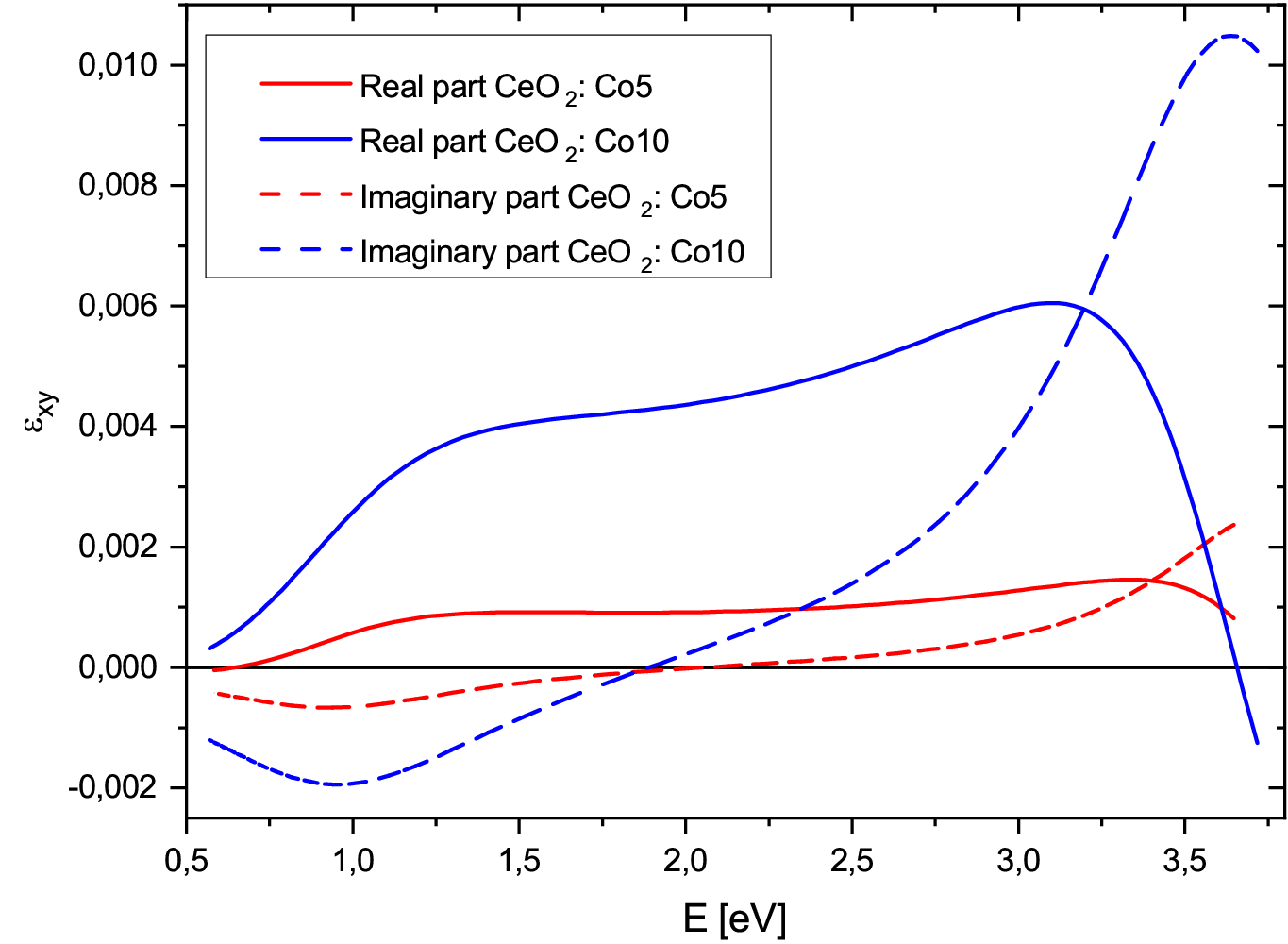}
	\caption{Spectra of the real and imaginary parts of the off-diagonal elements of the permittivity tensor of Ce$_{1-x}$Co$_x$O$_{2-\delta}$ films with $x=0.05$ and $0.10$, fitted from spectra of Faraday rotation and diagonal elements of the permittivity tensor.}
	\label{fig:OffDiagElements}
\end{figure}

In order to specify the electronic mechanism explaining the MO properties of Ce$_{1-x}$Co$_x$O$_{2-\delta}$ films, we parametrize the experimental FR spectra in terms of electronic transitions. Therefore, we used the FR spectra (displayed in Figure~\ref{fig:FRspectra}) together with the diagonal elements of the permittivity tensor (shown in Figure~\ref{fig:DiagElements}) to numerically calculate a model describing the off-diagonal elements of the permittivity tensor. The off-diagonal spectra were parametrized by a sum of two paramagnetic transitions with opposite signs, assuming their spectral dependence takes the form
\begin{equation}
	\varepsilon_{xy}=2\Gamma(\varepsilon^{''}_{xy})_{max}\frac{\omega(\omega^2-\omega^2_0+\Gamma^2)-i\Gamma(\omega^2+\omega^2_0-\Gamma^2)}{(\omega^2-\omega^2_0-\Gamma^2)^2+4\Gamma^2\omega^2},
\end{equation}
where transition strength $(\varepsilon^{''}_{xy})_{max}$, transition energy $\omega_0$, and damping constant $\Gamma$ were adjusted by least squares method. A summary of the oscillator parameters is given in Table~\ref{tab:TransitionParameters} and the resulting spectra of the real and imaginary parts of the off-diagonal elements of the permittivity tensor are shown in Figure~\ref{fig:OffDiagElements}. Assignment of the individual transitions is discussed below.

\begin{table}
	\caption{Parameters of two paramagnetic transitions describing the magneto-optical response of Ce$_{1-x}$Co$_x$O$_{2-\delta}$ films with $x=0.05$ and $0.10$; $(\varepsilon^{''}_{xy})_{max}$, $\omega_0$, and $\Gamma$ stand for the maximum amplitude, resonant frequency, and broadening of the oscillators. The parameters were fitted from the Faraday rotation spectra shown in Figure~\ref{fig:FRspectra} with use of the diagonal elements of the permittivity tensor presented in Figure~\ref{fig:DiagElements}.}
	\lineup
	\begin{indented}	
	\item[]\begin{tabular}{@{}lllllll}
	\br
	& & \centre{2}{Transition I: CF Co$^{2+}$} & & \centre{2}{Transition II: CT} \\
	& & \centre{2}{$^4A_2\rightarrow{}^4T_1(F)$} & & \centre{2}{Co 3$d\rightarrow$ Ce 4$f$} \\ \ns
	& & \crule{2} & & \crule{2} \\
	Sample & & CeO$_2$: Co5 & CeO$_2$: Co10 & & CeO$_2$: Co5 & CeO$_2$: Co10 \\ \mr
	$(\varepsilon^{''}_{xy})_{max}$ & & \-0.0009 & \-0.0036 & & 0.0026 & 0.0109 \\
	$\omega_0$ (eV) & & 0.90 & 0.87 & & 3.77 & 3.63 \\
	$\Gamma$ (eV) & & 0.5031 & 0.6133 & & 0.4182 & 0.5177 \\ \br 	
	\end{tabular}
	\end{indented}
	
	\label{tab:TransitionParameters}
\end{table}

One explanation of RT ferromagnetism in Ce$_{1-x}$Co$_x$O$_{2-\delta}$ would be a ferromagnetic coupling of cobalt ions. This was recently demonstrated for example by Tarui et al. \cite{Tarui2021} in Co-doped CeO$_2$ synthesized by a solid-state reaction method, however, the ferromagnetic response was due to formation of metallic precipitates of nanosized cobalt particles. At the low doping levels used in our study we expect the Co to be incorporated into the CeO$_2$ lattice without formation of metallic particles, as suggested by the lack of secondary phases in XRD analysis \cite{Bi2008}. The dilute Co content would produce only weak superexchange interactions.

Next, we address a potential contribution of oxygen vacancies, which is one of the most frequently discussed mechanisms in RT ferromagnetism of both the undoped and doped CeO$_2$. It is typically described as F$^+$-centre mediated exchange, in which a spin-polarized electron in an oxygen vacancy (F$^+$-centre) enables ferromagnetic exchange coupling between nearest-neighbouring cobalt ions, or Ce$^{3+}$ ions in case of undoped CeO$_2$ \cite{Ackland2018}. Although this mechanism is supported by some ab-initio calculations \cite{Song2009}, the strongest arguments for the involvement of oxygen vacancies in RT ferromagnetism are typically made in experimental reports, where various CeO$_2$ samples are treated in oxidizing or reducing conditions \cite{Wen2007,Singhal2010,Li2008,Singhal2012,Vodungbo2008,Song2010}, demonstrating even cyclical switching of the magnetization \cite{Shah2009}. This provides strong evidence in support of the F$^+$-centre mediated mechanisms. It is worth noting that after the first reduction and oxidation treatment the cyclical switching was completely reversible. However, it was not possible to recover the exact as-grown magnetization state, indicating possible presence of other defects or defect clusters. Therefore, additional mechanisms contributing to the overall magnetization of Ce$_{1-x}$Co$_x$O$_{2-\delta}$ cannot be ruled out.

Our work shows an increase in magnetization with Co content, similar to \cite{Yang2023} and in contrast with the inverse correlation reported in \cite{Song2007,Fernandes2011b}. Mahmoud et al. \cite{Mahmoud2015} have recently reported an increased magnetization with increased cobalt doping for $x\leq0.05$ for Ce$_{1-x}$Co$_x$O$_2$ films on LaAlO$_3$ made by a sol-gel process, but a decrease in magnetization for $0.05<x\leq0.15$, attributing the trend to a maximum in Co$^{2+}$/Co$^{3+}$, Co$^{3+}$/Co$^{4+}$ and oxygen vacancies at 5\% Co. As our films exhibit increased magnetization with Co content up to $x=0.25$ \cite{Bi2008}, we expect the defect chemistry in our Ce$_{1-x}$Co$_x$O$_{2-\delta}$ differs from that of \cite{Mahmoud2015}. The temperature dependence of the magnetic moment for $x=0.3$ and $0.5$ consists of two contributions \cite{Song2007,Tiwari2006} attributed to two different F$^+$-centre subsets \cite{Song2007}, or an electron transfer from the impurity bands to the empty $4f$ bands of Ce$^{4+}$ at higher temperatures \cite{Tiwari2006}. Since the magnetization \cite{Bi2008} together with the MO figure of merit in our films increases up to cobalt concentrations as large as $x=0.25$, we propose a mechanism contributing to the RT ferromagnetism in Ce$_{1-x}$Co$_x$O$_{2-\delta}$ that is directly related to cobalt ions.

Based on this hypothesis, the two paramagnetic transitions describing the off-diagonal elements of the permittivity tensor were assigned as follows. The first one is assumed to be a crystal field (CF) transition of Co$^{2+}$ between $^4$A$_2$ and $^4$T$_1(F)$ states, as inferred from studies of cobalt ferrites \cite{Fontijn1999}. Such transition is also supported by ab initio calculations in Co doped CeO$_2$, where splitting of the Co 3$d$ levels was reported at similar energies~\cite{Song2009,Shah2009,Ferrari2010}. The latter transition could be assigned as a charge transfer (CT) between O 2$p$ and Ce 4$f$ states, which was observed at similar energies in pure CeO$_2$ \cite{Barreca2003}. However, these transition energies were slightly higher compared to the second paramagnetic transition (see Table~\ref{tab:TransitionParameters}), as confirmed also by our ellipsometric data (cf. Figure~\ref{fig:DiagElements}), which showed the O 2$p$ to Ce 4$f$ at around 3.9~eV. Therefore, since ab initio studies on Co-doped CeO$_2$ usually report the presence of localized $d$ electrons of Co just above the O 2$p$ levels \cite{Ferrari2010,Song2009,Zimou2021}, we are more inclined to assign the second transition as a CT between Co 3$d$ and Ce 4$f$ states. Such assignment is in line with our suggested involvement of cobalt ions in the RT ferromagnetism in Ce$_{1-x}$Co$_x$O$_{2-\delta}$, and it also explains the slight energy difference ($\sim$200 meV) with respect to the most prominent optical CT transition from O 2$p$ states. The strength of the CF oscillator is significantly smaller compared to the CT one (see Table~\ref{tab:TransitionParameters} and Figure~\ref{fig:OffDiagElements}). The transition is parity forbidden and is observed only due to broken symmetry arising from tensile strain in the layer.

\begin{figure}[b]
	\centering
	\includegraphics[scale=0.5]{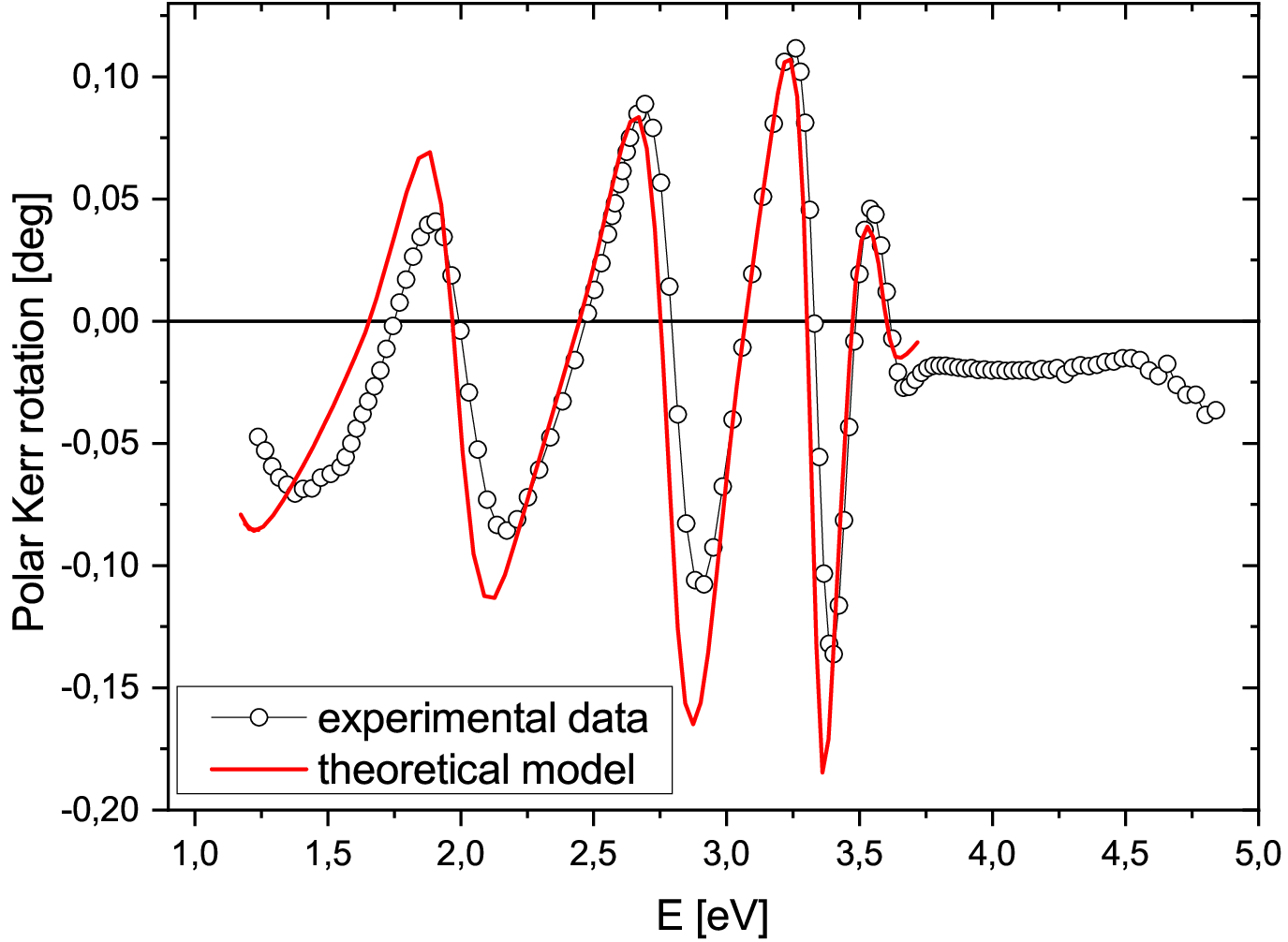}
	\caption{Comparison of experimental polar Kerr rotation spectrum of the Ce$_{0.90}$Co$_{0.10}$O$_{2-\delta}$ sample with a theoretical model. The experimental data were measured at room temperature and nearly normal light incidence. The theoretical data were calculated using the diagonal (Figure~\ref{fig:DiagElements}) and off-diagonal (Figure~\ref{fig:OffDiagElements}) elements of the permittivity tensor.}
	\label{fig:KerrSpectra}
\end{figure}

The assignment was further tested by LT measurements of Faraday effect, shown in the inset of Figure~\ref{fig:FRspectra}, and by MO Kerr effect measurements and calculations, Figure~\ref{fig:KerrSpectra}. The LT spectrum of FR, which was measured at 80~K, is essentially identical to the RT spectrum, both in amplitude and spectral behaviour. Negligible changes in the LT spectrum indicate no magnetic anisotropy changes with temperature and negligible strain variations with temperature. The RT spectrum of polar Kerr rotation, measured at nearly normal light incidence, was compared to theoretical calculations based on the diagonal elements of permittivity tensor shown in Figure~\ref{fig:DiagElements}, and the off-diagonal elements parametrized by the two paramagnetic oscillators (see Table~\ref{tab:TransitionParameters} and Figure~\ref{fig:OffDiagElements}). An excellent agreement between experimental and theoretical data is clearly visible, confirming proper spectral dependence of the obtained elements of the permittivity tensor. Therefore, both the LT Faraday effect measurements and the Kerr effect measurements further supported the proposed theoretical model, formed by two paramagnetic transitions involving cobalt ions.

\section{Conclusion}

Optical and MO properties of Ce$_{1-x}$Co$_x$O$_{2-\delta}$ thin films grown on MgO ($x=0.05$ and $0.10$) and oxidized Si ($x=0.20$) were reported in a spectral range from near IR to UV. The complete permittivity-tensor spectra were provided for further use in prospective modelling of new devices. According to the experimental data, cobalt doping has a significant influence on both optical and MO properties of the samples. Unlike several previous studies, we find that ferromagnetic properties are enhanced with increasing cobalt concentration up to $x=0.20$. We also find little evidence for oxygen vacancies in the optical spectra. Therefore, we propose a theoretical model of two paramagnetic transitions involving cobalt ions to describe the MO properties, one a crystal field transition of $Co^{2+}$ and the other a charge transfer between Co 3$d$ and Ce 4$f$ states. This model is in agreement with both the RT and LT Faraday effect spectroscopy, and it is further supported by an agreement of experimental and theoretical spectra of MO Kerr effect. We therefore conclude that the major influence of cobalt doping on MO behaviour of Ce$_{1-x}$Co$_x$O$_{2-\delta}$ might be due to CF and CT transitions involving cobalt ions.

\ack
M.Z. acknowledges the support of the THRILL (Grant Agreement No. 101095207) project from the Horizon Europe of the European Union. M.V. acknowledges MATFUN (Grant No. CZ.02.1.01/0.0/0.0/15\_003/0000487). M.V. and C.A.R. thank G.F. Dionne for very fruitful discussions about electronic structure and optical properties of the investigated material.

\clearpage

\section*{References}
\bibliographystyle{iopart-num}
\bibliography{bibliography}

\end{document}